\documentclass[sigconf]{acmart}
\renewcommand\footnotetextcopyrightpermission[1]{} 
\usepackage{fancyhdr}
\usepackage{color}
\usepackage{diagbox}
\usepackage[linesnumbered,ruled]{algorithm2e}
\usepackage{amsmath}
\usepackage{subfigure}
\usepackage{nicematrix}
\usepackage{multirow}

\newcommand{\code}[1]{{\small\texttt{#1}}}

\newcommand{\para}[1]{\medskip\noindent\textbf{#1}}

\begin{document}
\title{PPCA: Privacy-preserving Principal Component Analysis Using Secure Multiparty Computation (MPC)}

\author{Xiaoyu Fan}
\email{fanxy20@mails.tsinghua.edu.cn}
\affiliation{%
  \institution{IIIS Tsinghua University}
  \city{Beijing}
  \country{China}
}
\author{Guosai Wang}
\email{wanggs@pbcsf.tsinghua.edu.cn}
\affiliation{%
  \institution{PBC School of Finance, Tsinghua University}
  \city{Beijing}
  \country{China}
}
\author{Kun Chen}
\email{chenk@pbcsf.tsinghua.edu.cn}
\affiliation{%
  \institution{PBC School of Finance, Tsinghua University}
  \city{Beijing}
  \country{China}
}
\author{Xu He}
\email{Larryhe@tsingj.com}
\affiliation{%
  \institution{Huakong Tsingjiao Information Technology Inc.}
  \city{Beijing}
  \country{China}
}
\author{Wei Xu}
\email{weixu@tsinghua.edu.cn}
\affiliation{%
  \institution{IIIS Tsinghua University}
  \city{Beijing}
  \country{China}
}

\begin{abstract}
Privacy-preserving data mining has become an important topic. People have built several multi-party-computation (MPC)-based frameworks to provide theoretically guaranteed privacy, the poor performance of real-world algorithms have always been a challenge.  
Using Principal Component Analysis (PCA) as an example, we show that by considering the unique performance characters of the MPC platform, we can design highly effective algorithm-level optimizations, such as replacing expensive operators and batching up. We achieve about 200$\times$ performance boost over existing privacy-preserving PCA algorithms with the same level of privacy guarantee.  Also, using real-world datasets, we show that by combining multi-party data, we can achieve better training results.  

\end{abstract}



\maketitle

\section{Introduction}

Mining datasets distributed across many parties without leaking extra private information has become an important topic recently.  Integrating data from multiple parties increases the overall training dataset and provides an opportunity to train on datasets with different distributions and dimensions.  In fact, researchers have shown that by integrating data from multiple parties, we can improve model prediction accuracy~\cite{zheng2020industrial}, and even train models that we have not been able to~\cite{10.1145/543613.543644}. 

However, privacy concerns have been a big hurdle in data integration.  Privacy-preserving algorithms offer some promising solutions.  
Private information to protect goes beyond personal data (a.k.a. personal identifiable information, or PII) and includes other less obvious private information, such as data distribution,  as adversaries may infer important business information from it.  For example, ~\cite{NEURIPS2019_60a6c400, zhao2020idlg} show attacks in which the adversaries can obtain the information and defeat the purpose of privacy protection, just by analyzing the intermediate results without seeing the original data.  Note that these leaks depend on the data and model, and their security properties are sometimes hard to reason about.

Secure Multi-Party-Computation (MPC)~\cite{4568207} is a class of cryptographic techniques that allow multiple parties to jointly compute on their data without leaking any information other than the final results.  Although it provides a theoretically promising solution, its performance is far from useful in analyzing large-scale datasets even using simple data mining algorithms.  

Recently, people have proposed MPC frameworks that greatly improve the computation efficiency~\cite{mohassel2017secureml, demmler2015aby, li2019privpy, ryffel2018generic}.  Most importantly, these frameworks can theoretically support any algorithm on MPC by offering high-level programming languages, such as Python~\cite{li2019privpy, ryffel2018generic}.  However, performance remains a challenge for sophisticated algorithms, even for small datasets.  

We believe there are three main challenges to solve the problem: 

Firstly, these MPC frameworks mainly focus on the performance of basic operations such as multiplications and comparisons, but the performance of data mining algorithms depends on complex operations such as division and square-root~\cite{pathak2010privacy, vaidya2003privacy, han2008privacy}.  
In MPC, multiplication and comparison require communications among the parties, so they are slower than addition.  
Also, people use numerical methods (such as Newton's method) to implement complex operations, and thus the time cost can be orders-of-magnitude higher than multiplication.  
Table~\ref{Table:time costs} in Section~\ref{section:MPC_features} shows the comparison among different operations in an MPC system.  

Secondly, using the provided high-level language, the naive way to implement algorithms is to directly port the algorithm designed for plain-text onto MPC platforms.  The naive porting often leads to poor performance due to the huge performance gap among different basic operations.  Also, there may be some asymptotic complexity differences in algorithms like sorting~\cite{jonsson2011secure}, as we have to hide the comparison results among elements.  

Last but not least, many of these frameworks rely on data partition and parallelism to provide large-scale data processing in a reasonable time, but some data mining algorithms, such as PCA, are non-trivial to partition into independent tasks.  

In this paper, using the \textit{principal component analysis (PCA)} as an example, we demonstrate an end-to-end optimization of a data mining algorithm to run on the MPC framework.  We choose to use PCA not only because it is a popular algorithm but also because it involves several steps that require a variety of performance optimization techniques to illustrate an end-to-end data mining method.  Although we implemented our algorithm on a single MPC framework, the optimization techniques and tradeoffs are independent of the specific framework. 

The first step of PCA is to calculate the covariance matrix from raw data from the participating parties.  The dataset can be horizontally partitioned (i.e., each party has the same feature space with different samples) or vertically partitioned (i.e., each party has the same sample space with different features).  As the raw data can be large, it is important to avoid too many computations on cipher-text.  We apply some transformations in covariance matrix computation to avoid encrypting all raw data by locally computing the partial results and only aggregate the partial results using MPC. 

The next step is the core step of PCA, in which we perform eigen-decomposition on the covariance matrix.  Previous work~\cite{al2017privacy} provides an MPC-based approach to compute it on cipher-text, but the performance is unacceptable (even a $50 \times 50$ matrix takes 126.7 minutes). The slow performance leads to proposals to reveal the covariance matrix as plain-text for decomposition, arguing that it does not contain private information~\cite{liu2020privacy}.  However, as our analysis in Section~\ref{sec:problem_statement} shows, the covariance matrix does leak much information about the distribution of each party's dataset in many cases. Also, we need to sort on the eigenvalues to select the largest $K$ corresponding eigenvectors to construct the projection matrix. We carefully designed the decomposition and sort algorithms based on the MPC platform's performance characters, avoiding expensive operations and fully exploiting the opportunities to batch up operations, thus greatly reduced the computation time.  In our work, we have improved performance by $200 \times$ comparing with~\cite{al2017privacy} on similar scale matrices and have achieved the entire PCA on the $7,062,606 \times 115$ dataset from 9 parties within 3 minutes.  


Using real-world datasets, we show that performing PCA from integrated datasets from multiple parties can benefit downstream tasks. We have adopted two real-world datasets for demonstration in Section~\ref{Section:exp-effect}.

In summary, there are four contributions of this paper:
\begin{itemize}
    \item We designed an end-to-end privacy-preserving PCA algorithm implementation optimized for MPC frameworks; 
    \item We proposed an optimized eigen-decomposition algorithm and sort algorithm fully taking into account the basic operation cost of MPC algorithms;
    \item We demonstrated that we could perform the entire PCA algorithm on a $7,062,606 \times 115$ dataset within 3 minutes.  
    \item Using real-world datasets, we show that integrating data from different parties can significantly improve the downstream models' quality.
\end{itemize}
\section{Related Work}

We first compare different underlying techniques that achieve general privacy-preserving data mining and then review the studies on privacy-preserving PCA algorithms. 

\para{Privacy-preserving techniques.}
There are four popular lines to achieve privacy-preserving data mining:
\emph{MPC}, \emph{differential privacy (DP)}~\cite{dwork2014algorithmic}, \emph{federated learning (FL)}~\cite{yang2019federated}, and \emph{trusted execution environment (TEE)}~\cite{7345265}.
DP-based techniques introduce random noises to defend against differential queries or attacks, but the side effect is the precision loss of the computation results~\cite{dwork2014algorithmic, 10.1145/1835804.1835868}.
FL exchanges intermediate results such as the gradients instead of the raw data among multiple data providers to reduce privacy leakage when training models~\cite{yang2019federated}. 
For PCA, FL methods have to exchange parts of the raw data to compute the covariance matrix, and expose the matrix to decompose it, breaking the privacy requirement.
TEE runs algorithms in an isolated secure area (a.k.a. enclave) in the processor~\cite{7345265}, but its security depends on the hardware manufacturer, and sometimes is vulnerable to side-channel attacks~\cite{shih2017t}.  

MPC achieves secure computation using cryptographic techniques while performance remains a challenge. There are many MPC frameworks trying to improve basic MPC operation performance. However, we still need algorithm-level optimizations to make algorithms like PCA practical on large datasets.

\para{Privacy-preserving algorithms.}
Researchers have studied various privacy-preserving data mining algorithms. Including privacy-preserving classification~\cite{du2004privacy, chabanne2017privacy, yang2005privacy}, regression~\cite{nikolaenko2013privacy, sanil2004privacy} and unsupervised algorithms like K-Means\cite{jagannathan2005privacy} and EM clustering~\cite{lin2005privacy}.

People have provided several privacy-preserving PCA designs too.  \cite{8456056} uses a well-designed perturbation to the original data's covariance matrix to keep it secret.  In addition to the inaccurate results, it does not provide a way to jointly compute the covariance matrix. \cite{liu2020privacy} supports joining horizontally-partitioned raw data, but it exposes the covariance matrix to perform the eigen-decomposition.  As we will show in Section~\ref{sec:problem_statement}, the covariance matrix may leak important information about each party's raw data. \cite{al2017privacy} handles horizontally-partitioned dataset and allows keeping the covariance matrix private.  However, it takes $127.4$ minutes even on a small dataset with only $50$ features.  We adopt the pre-computation techniques in Section~\ref{section:covariance construction} for the horizontally partitioned data, but provide orders-of-magnitude performance improvement.

Some studies focus on privacy-preserving matrix decomposition, the slowest step in PCA~\cite{pathak2010privacy, 4812517}.  
\cite{pathak2010privacy} uses an MPC-based \textit{power iteration} method to calculate only the larggest eigenvalue and eigenvector.  
\cite{4812517} uses the QR algorithm to implement a privacy-preserving SVD algorithm of input matrix $M$ that can be written in the form $BB^{T}$ or $B^{T}B$ where $B$ is some matrix that is horizontally or vertically partitioned among exactly two parties. Also, it takes over $166.67$ minutes to decompose a $50 \times 50$ matrix.

Compared to previous studies, our method offers the following desirable properties:  
1) keeping everything private, including the covariance matrix.
2) supporting any number of parties that hold the partitioned data,
3) supporting both horizontally and vertically partitioned dataset, and
4) running fast and outputting all the eigenvalues and eigenvectors.

\section{Background}
\label{sec:background}

In this section, we provide some background information. Specifically, we summarize important performance characteristics of MPC protocols, which leads to our optimization design.

\subsection{Secure Multi-Party Computation (MPC)}
\label{section:MPC protocols}

MPC has a long history in the cryptography community.  It enables a group of data owners to jointly compute a function without disclosing their data input.  

\para{Basic MPC protocols.  }
To achieve the generality of MPC, i.e., to support arbitrary functions, people usually start by designing basic operations using cryptographic protocols. 

There are protocols focusing on supporting a single important operation.  
For example, \emph{oblivious transfer (OT)} protocol allows a receiver $R$ to select one element from an array that the sender $S$ holds, without letting $S$ know which one she selects~\cite{goldreich2019proofs}.  
\emph{Secure multi-party shuffling (MPS)}~\cite{movahedi2015secure} allows parties to jointly permute their inputs without disclosing the input itself. The protocol is useful to implement functions like  \code{argsort}.

There are also protocols allowing general MPC.  
\emph{Secret sharing (SS)} protocol allows splitting a secret input $s \in \mathbb{R}$ into $n$ pieces $(s_1, s_2, ..., s_n)$, and let each computation server to hold one of the pieces~\cite{beimel1993interaction}.  We can perform basic operation protocols over these shared secrets, such as multiplication and comparison, without allowing any of the parties to see plain-text data, under certain security assumptions~\cite{10.1007/3-540-48329-2_12}. 
\textit{Garbled Circuits (GC)}~\cite{yao1982protocols} treats the function to compute as a look-up table, and each party's input as keys addressing into the table using protocols like oblivious transfer.  GC is proven to be general to any computation; however, its efficiency and scalability to a large number of parties remain a challenge.  Thus, most practical frameworks are based on SS~\cite{chen2020secure, 10.1145/3394138}.

\para{Composed protocols.  }
Although we can perform any computation using the MPC protocols like SS or GC in theory, one obstacle is that designing a more complex operation directly based on these protocols are both difficult to program and inefficient.  Thus, most practical MPC frameworks only implement basic operations using these protocols, and then \emph{compose} them to implement more complex functions~\cite{mohassel2017secureml, li2019privpy}, just like running programs on an instruction set of a processor.  The security guarantee of composing these operations are not obvious, but many of them are proven to be composable in cryptography and adopted in existing MPC frameworks~\cite{demmler2015aby, mohassel2017secureml, li2019privpy}.  

Obviously, all these operations are slower than the plain-text version, mainly because 1) the data size in cipher-text is larger than plain-text; and 2) even basic operations, such as multiplication, involves one or more rounds of communications among the computation nodes~\cite{demmler2015aby, damgaard2012multiparty, li2019privpy}.  These optimizations usually focus on single operations but not complex algorithms. 

\para{Practical MPC frameworks.  }
As the computation involves multiple parties, people have built practical MPC frameworks to reduce the algorithm development and system management overhead.  These frameworks provide a programming interface, allowing users to specify data mining algorithms on MPC using a high-level language for the computation platform to execute the composed protocols.  A typical computation engine includes 1) modules that run at each data owner party to encrypt the data; 2) a set of MPC servers to execute the MPC protocols, and 3) a module to decrypt the results in the designated receiving party.   

Many frameworks improve basic operation performance (throughput or latency, or both) by leveraging modern technology such as multi-core processors, dedicated hardware accelerators~\cite{omondi2006fpga}, low latency networks, as well as programming models like data and task parallelism.  

As an example, PrivPy~\cite{li2019privpy} is a MPC framework designed for data mining. It is based on the \textit{2-out-of-4 secret sharing} protocol, and it offers a Python programming interface with high-level data types like NumPy~\cite{oliphant2006guide}. It treats each joint party as a client. Each client sends a secretly shared data to four calculation servers. Then the servers perform the SS-based privacy-preserving algorithms.

\subsection{Operation Characteristics of MPC}
\label{section:MPC_features}

\begin{table}[t] 
	\small
	\caption{Time cost of basic operations}\label{Table:time costs}
	\begin{tabular}{ccccccccl}
		\toprule
		Input size & \code{add} & \code{mul} & \code{gt} & \code{eq} & \code{sqrt} & \code{reciprocal} \\
		\midrule
		$10^4$ & $1/50$ & $1$ & $10$ & $20$ & $80$ & $90$ \\
		$10^5$ & $2/50$ & $4$ & $20$ & $60$ & $240$ & $270$ \\
		$10^6$ & $13/50$ & $9$ & $80$ & $180$ & $1600$ & $1350$ \\
		\bottomrule
	\end{tabular}
\end{table}

The cipher-to-plain-text performance difference is still large, and several MPC frameworks are trying to improve it.  In this paper, we only focus on the relative performance characteristics of different MPC operations that are not likely to change with framework optimizations yet have a big impact on algorithm design.

Using~\cite{li2019privpy} as an example, we compare the \textit{relative time} for each secure operations useful in PCA, normalizing to the time taken to compute 10K multiplications (MULs). Table~\ref{Table:time costs} shows a summary.  Actually, the relations among each operation are general across many MPC frameworks, and we will explain the main characteristics in the following.





1) \emph{Addition is almost free comparing to MUL, while comparison is over 10$\times$ more expensive.}  This is because all compelling frameworks use protocols that avoid communication in addition, and only a single round of communication for MUL. For comparison, however, we need to perform many rounds of bit-wise operations, and the exact number depends on the design protocol. Normally 10+ rounds are required for practical MPC frameworks~\cite{li2019privpy, bogdanov2008sharemind, nishide2007multiparty}.

2) \emph{Non-linear mathematical functions are close to 100$\times$ slower than MUL. }
Numerical method is a typical way to implement these functions.  For example, We use \emph{Newton approximation} to compute \code{reci} (reciprocal) and \code{sqrt} (square root), and we illustrate the algorithms in Appendix~\ref{sqrt and reciprocal}.  One observation is that these numerical methods require multiple iterations. As the input is in cipher-text, many early-termination optimizations in plain-text no longer apply, and thus these operations are much slower than MUL. 

3) \emph{The speed-up from batch processing is significant. }
As most operations require communications, there is a fixed overhead to establish the connection, encode the data and generate keys. Thus it is important to batch-up operations to amortize this overhead.  We observe the same speed-up of batch processing, as reported in~\cite{damgaard2012multiparty, mohassel2017secureml, demmler2015aby, li2019privpy}.  For example, in Table~\ref{Table:time costs} we observe a speed-up of $9\times$ for EQ when we increase input batch size from $10^4$ to $10^6$. 

These characteristics we observe in MPC are apparently different from the relative performance in plain-text, which leads to a different set of algorithm design choices. 

\section{Privacy-Preserving PCA Design}
  
We introduce the optimizations on the PCA algorithm in this section.
We first formalize the privacy-preserving PCA's objective, and then describe the detailed design and highlight our optimizations.

\subsection{Problem Definition}
\label{sec:problem_statement}

A group of $N$ \emph{data providers} $p_1, p_2, \dots, p_N$ jointly hold the \emph{original dataset}, a matrix $X$ of size $n\times d$ ($n$ samples of $d$ features). 
Each data provider $p_i$ holds some rows or some columns of $X$, represented by $X_i$.
The group's target is to collaboratively perform a \textit{principal component analysis (PCA)} and compute the \emph{projection matrix} $P$ of size $d \times K$, where $1 \le K \le n$ is an input to the algorithm, such that the $i$-th column of $P$ is the eigenvector corresponding to the $i$-th largest eigenvalues of the covariance matrix $C=\frac 1 {n-1} \overline{X}^T \overline{X}$, where $\overline{X_{i}} = X_{i} - \mu_{i}, 1 \leq i \leq d$.

Each party wants to protect her own original data, and meet the following requirements: 1) 
No other party $p_k (k \neq i)$ and any computation server $E$ can infer the privacy of data $X_i$ held by $p_i$ during the algorithm execution; 2) No party $p_i$ and any computation server can infer the information about the covariance matrix $C$; and 3)  $P$ is a correct projection matrix of $X$ under normal PCA definition. Besides, the algorithm needs to run fast on a modern MPC framework.  These guarantees should hold with no extra security assumptions other than those required by the underlying MPC framework. 

We emphasis that requirement 2) is necessary to achieve 1).  Previous work~\cite{liu2020privacy} proposes to reveal $C$ to accelerate the computation.  We show that a plain-text $C$ can cause leaking information of $X$.  Given the covariance matrix $C \in \mathbb{R}^{d \times d}$:
\begin{equation}
\begin{aligned}
C_{i, j} = \frac{1}{n-1}\sum_{k=1}^{n}{(x_{jk} - \overline{x_j})(x_{ik} - \overline{x_i})}, \quad i, j = 1, 2...,d, 
\end{aligned}
\end{equation}
where $i = j, C_{i, i}$ is the variance of the $i$-th feature.   Using a gradient descent with the objective function
\begin{equation}
\min_{X} \quad J = ||X^{T}X - (n-1)^2C||_{F}^{2}\, ,
\end{equation}
we can obtain a number of candidate $X'$s, all of them would produce the same covariance matrix $C$.  When other information are available and depending on the distribution of the original dataset $X$, an adversary can sometimes find out a $X'$ that is very close to $X$, causing a failure to meet requirement 1).  

We highlight the reason why we choose privacy-preserving PCA as an example to illustrate our optimization framework. 1) The PCA process contains many expensive steps like eigen-decomposition and sort, 2) local preprocess is an important optimization in both scalability and efficiency, 3) the optimized building blocks like eigendecomposition and sort are common in many other applications.

\subsection{Method Overview}

\begin{figure*}[tb]
	\centering
	\includegraphics[scale=0.28]{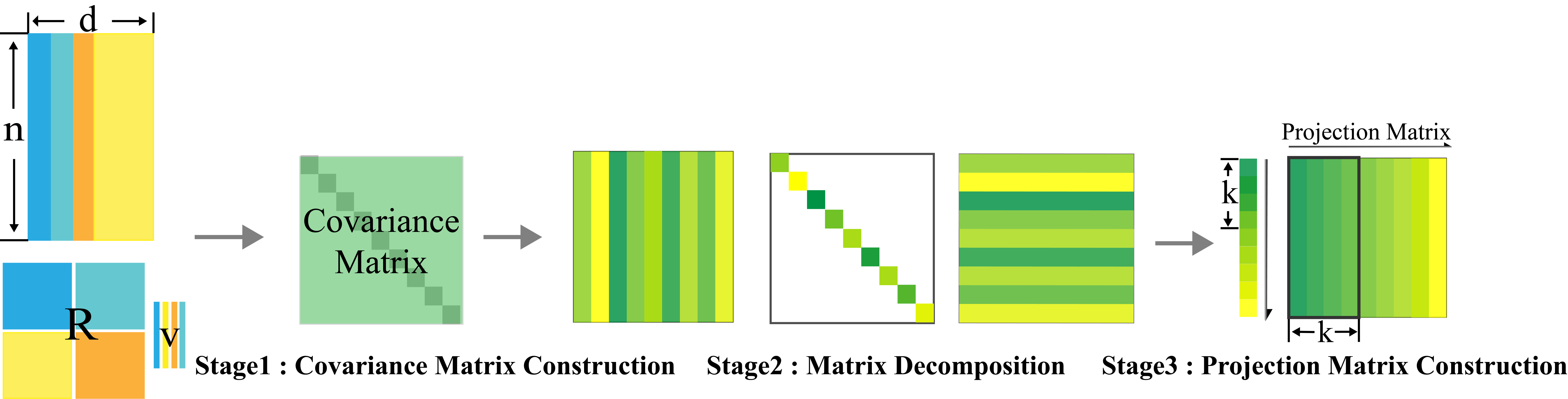}
	\caption{End-to-end Privacy-preserving PCA's framework}\label{Figure:cipher calculation}
\end{figure*}

Figure~\ref{Figure:cipher calculation} illustrates the four major steps.

\para{Step 1) Covariance matrix construction. } To reduce the amount of computation on cipher-text, we first let each party preprocess the data before she encrypts and sends out the data to compute the covariance matrix in the MPC framework.  The computation depends on how the dataset is partitioned at each party, and we support both horizontal and vertical partitions.  


\para{Step 2) Eigen-decomposition.  } This is the most time consuming step.  We use an optimized Jacobi method to perform the computation and avoid expensive operations on MPC framework. 
	
\para{Step 3) Projection matrix construction.  } We need to choose the largest $K$ eigenvalues. To sort the eigenvalues fast, we developed a $batch\_sort$ algorithm that significantly improved efficiency through batch operations. 

\para{Step 4) Inference.  }  Optionally, we can keep the projection matrix in cipher-text, and perform downstream tasks like dimension reduction and anomaly detection without decrypting the data. 
  

  
\subsection{Covariance Matrix Construction}\label{section:covariance construction}

\begin{figure}[h]
\centering
\subfigure[Horizational]{
  \includegraphics[scale=0.28]{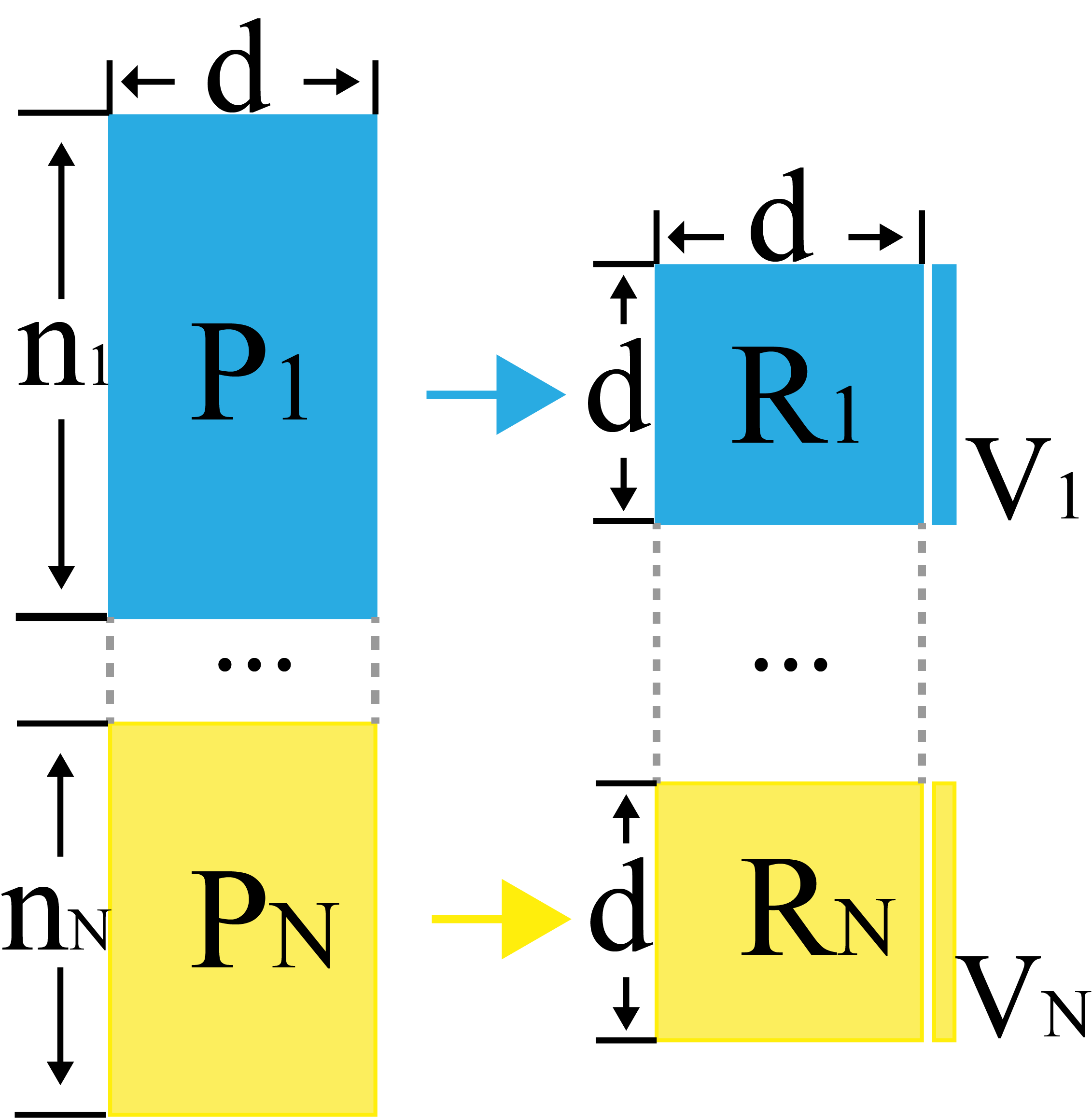}
}\hspace{7mm}
\subfigure[Vertical]{
  \includegraphics[scale=0.28]{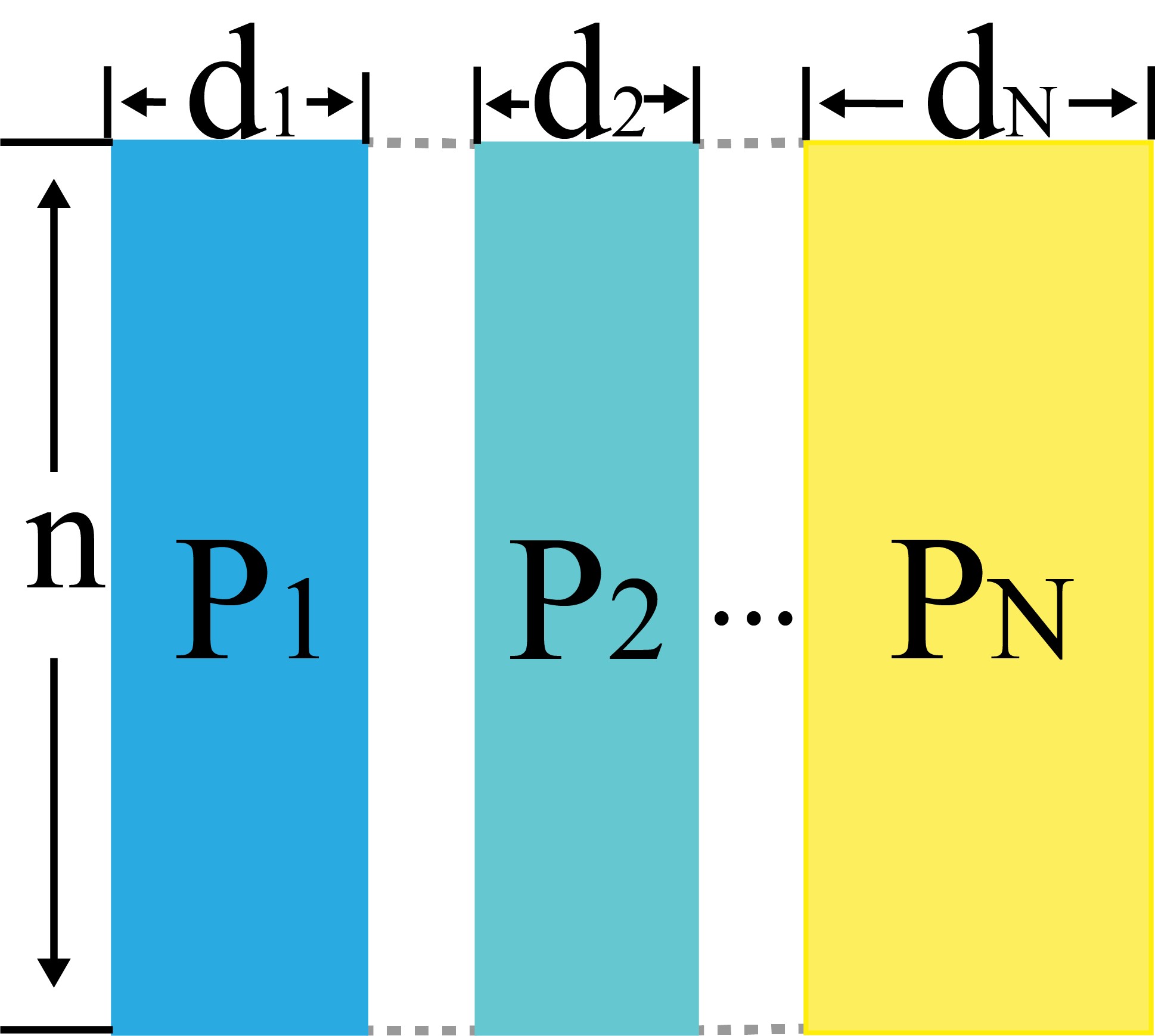}
}
\caption{Data partitions and Local preprocess}\label{Figure:locally preprocess}
\end{figure}

In the first step, we perform local preprocessing on each data owner, and jointly compute the covariance matrix, depending on the partation of the dataset.

\para{Horizationally Partitioned Datasets.}
\label{section:hp}
In this common case, each \emph{data provider} $p_i$ owns a subset of rows (i.e. samples) of the origional dataset $X$, i.e.,
\begin{equation}
\begin{aligned}
X & = 
\begin{bNiceMatrix} 
X_{1}, 
X_{2}, 
...,
X_{N}
\end{bNiceMatrix}^{T} \in \mathbb{R}^{n \times d} ,\\
\end{aligned} 
\end{equation}
where $X_{i} \in \mathbb{R}^{n_{i} \times d}$ is the subset of rows of $p_i$ of size $n_i$.
The key optimization is to allow each party process as many samples as possible locally to avoid expensive MPC operations for covariance matrix computation. For $N$ parties, we let each party $p_i$ locally compute partial results $R_{i} = \sum_{j=1}^{n_i}{x_{j}x_{j}^{T}}$ and $v_{i} = \sum_{j=1}^{n_i}{x_{j}}$. 
Then we can jointly compute the covariance matrix of $X$, $C$, using the same method as~\cite{al2017privacy}:
  \begin{equation}
      C = \frac{1}{n - 1} \sum_{j=1}^{N}R_{i} - \frac{1}{n(n-1)}vv^{T} , 
  \end{equation}\label{eq:horizontal-preprocess}
where $v = \sum_{i=1}^{N}v_{i}$ and $n = \sum_{i=1}^{N}{n_{i}}$. Note that the data we need to compute on MPC platform for each party $p_i$ reduces from $n_{i} \times d$ (i.e., the dataset $X_i$) to $(d \times d + d)$ (i.e., $R_i$ and $v_i$), making the MPC complexity independent of the sample size. 


\para{Vertically Partitioned Datasets. }
\label{section:vp}
Each $p_i$ owns some columns (or features) of the dataset $X$.  Typically a common \emph{id} exists to align the data rows.   As the covariance matrix computation requires the operations using different columns, we need to perform it on the MPC platform.  If the \emph{id} is from a small namespace of size $\gamma$ (i.e. $\gamma$ is close to number of samples $n$), we map each sample to the namespace, filling missing samples with zeros.  Thus, each $X_i$ has dimension of $\gamma \times d_i$.  In the cases when $\gamma >> n$, we use a common cryptography protocol \emph{private set intersection (PSI)}~\cite{dong2013private} to compute the joint dataset $X$.

In either case, in order to scale the computation to support $\gamma$ or $n$ samples, we use data parallelism by partitioning the dataset into $M$ separate pieces, and perform an MPC task independently on each piece before merging the result as
\begin{equation}\label{eq:vertically fused data}
\begin{aligned}
C = X^TX = \sum_{j=1}^{M} {X_{j}^{T}X_{j}}, 
\end{aligned}
\end{equation}
where each piece $X_{j}$ has shape of $\frac{n}{M} \times d$, $\frac{n}{M} \in \mathbb{Z}$.

\subsection{Eigen-decomposition}\label{section:md}
  The eigen-decomposition is the most time-consuming step for PCA in both plain-text and cipher-text. In this section, we introduce how we select the appropriate algorithm prototype and how to optimize it based on the characteristics of MPC frameworks.
  

Eigen-decomposition is a well-studied topic.  There are three popular designs: \emph{Power Iterations}, \emph{QR shift} and \emph{Jacobi's method}.  \emph{Power Iteration}~\cite{al2017privacy, pathak2010privacy} is efficient, but it does not find all eigenvalues and eigenvectors and thus not suitable in our case.
QR shift~\cite{greenbaum1989experiments, rutter1994serial}, or QR decomposition with element shift for acceleration, based on \textit{householder reflection}, is one of the most popular methods in the plain-text implementations, and previous privacy-preserving algorithms adopted this method~\cite{4812517, grammenos2020federated}.
Jacobi method, with a parallel version, \emph{Parallel Jacobi}~\cite{10.2307/2005221}, is based on \textit{givens rotation}.  It is promising because it allows vectorizing many operations, thus may be suitable in many vector-operation-friendly MPC platforms.  We evaluate the latter two methods in this paper. 

In both algorithms, the operations that caused the huge computational overhead is the iteratively orthogonal transformations, i.e., \textit{Householder Reflection} and \textit{Givens Rotation}. Appendix~\ref{orthogonal transformation} shows the details of their computation. We introduce the privacy-preserving QR Shift and Vectorized Jacobi algorithm and show how to select the better algorithm in cipher-text next.


\para{QR Shift  } is based on the \textit{householder reflection (HR)} orthogonal transformations that finds an orthogonal matrix $H = I - 2ww^{T}$.  $H$ reflects the vector $w \in \mathbb{R}^{n}$ to $\mathbf{u} \in \mathbb{R}^{n}$, where $u_0 = ||w||_{2}$ and $u_{i} = 0$, for $0 < i < n$.

To compute QR Shift on a $d \times d$ matrix, we first apply $(d-2)$ HRs to reduce the original matrix to tridiagonal and then perform QR transformations iteratively until it converges to diagonal.  Each QR transformation requires $(d-1)$ HRs with a \textit{Rayleigh-quotient-shift}. Appendix~\ref{pp QR shift} shows the algorithm for both plain-text and MPC frameworks.

A variation of QR shift, the \emph{DP QR shift} uses divide-and-conquer to scale up the computation and thus popular on the plain-text platform.  However, it requires very expensive operations on MPCs, such as full permutation and solving \textit{secular equations}.  Thus, we do not consider this DP approach. 


\para{Parallel Jacobi } is based on \textit{givens rotation (GR)} orthogonal transformations that finds the orthogonal matrix $Q_{kl}(\theta) = \{q_{ij}(\theta)\}_{i, j=1}^{n}, \\1\leq k < l \leq d$, where $q_{kk}(\theta) = q_{ll}(\theta) = \cos(\theta), q_{kl}=-\sin(\theta), q_{lk} = \sin(\theta), q_{ii}=1$ for $i \neq k, l $ with other elements all zero. For any real-value symmetric matrix $A = \{a_{ij}\}_{i, j=1}^{n} \in \mathbb{R}^{d \times d}$, we can transform the off-diagonal elememt $a_{kl}, k < l$ to zero by rotation angle $\theta$ with the cotangent value statisfying  
  \begin{equation}
    \cot(2\theta) = \frac{|a_{kk} - a_{ll}|}{2*a_{kl}}. 
  \end{equation}

\textit{Jacobi's Method} produces a sequence $A_{i}, i = 0, 1, ...,$ of orthogonally similar matrices for a given real symmetric matrix $A = A_{0}$ that converges to a diagonal matrix.  We then compute the next matrix $A_{i+1}$ iteratively with
      \begin{equation}
        A_{i+1} = Q_{i}^{T} A_{i} Q_{i},
      \end{equation} 
where $Q_{i}$ is the orthogonal matrix determined using GR with the largest off-diagonal element of $A_{i}$.

Given the largest off-diagonal element $a_{kl}$, the rotation only affects the $k$-th and $l$-th rows and columns. 
There is an inherent parallelism property of Jacobi's method~\cite{10.2307/2005221}:  at most $\lfloor \frac{n}{2} \rfloor$ off-diagonal elements in the upper triangle can be eliminated to zero together in each round.  
There are two conditions to select elements in each round to improve the performance. We illustrate the selection strategy with the following $7 \times 7$ matrix as an example. Elements with same number are in the same batch for one iteration. 
     $$
      \begin{bNiceMatrix}
      \small
        a_{0, 0} &   &  &   &  &  &  \\
        3 &a_{1, 1}   &  &  &  & &7\\
        6 & 2 & a_{2, 2} & &  & 7 & 3\\
        2& 5 & 1 &a_{3, 3} & 7 & 3 & 6\\
        5 &1 &4 &  & a_{4, 4} &6 & 2\\
        1 & 4 &  &  & & a_{5, 5} & 5\\
        4 &  &  &  &  &  &a_{6, 6}\\
      \end{bNiceMatrix}.
      $$
We can aggregate the computation of the rotation matrix for a batch of elements. Here we use the first batch as an example, where $c_{i,j}$ and $s_{i,j}$ refers to $\cos\theta_{i,j}$ and $\sin\theta_{i,j}$ for the elements numbered 1.
      \begin{equation*}
      \small
        Q_{1} = Q_{1}^{1}Q_{1}^{2}Q_{1}^{3} = \\ 
        \begin{bNiceMatrix}
          c_{0, 5} & 0& 0 &0& 0&-s_{0, 5}&0 \\
          0 & c_{1, 4} & 0 &0  & -s_{1, 4} & 0& 0\\
          0 & 0 & c_{2, 3} & -s_{2, 3} & 0 & 0 & 0\\
          0 & 0 & s_{2, 3}  & c_{2, 3}  & 0 & 0 & 0\\
          0 & s_{1, 4} & 0 &0 &c_{1, 4} & 0 & 0 \\
          s_{0, 5} & 0& 0 & 0 &0 & c_{0, 5} & 0 \\
          0 & 0 & 0 &0 &0 & 0 & 1 \\
        \end{bNiceMatrix}.
      \end{equation*}

We leverage the parallel feature of Jacobi to implement independent operations in each iteration as a single vector operation, illustrated in Algorithm~\ref{Algo:vectorized jacobi}.   
Specifically, we first extract the independent elements to an array, and then calculate the rotation angles as a vector operation (Line 8).  The algorithm converges for any symmetric matrix, as there is a strong bound $S(A') \leq cS(A)$ where $c < 1, S(A) = \frac{\sqrt{2}}{2}||A - diag(A)||_{F}$ and $A'$ is the transformation of $A$ after updating all off-diagonal elements~\cite{hari2016convergence}.


      \begin{algorithm}[b]
        \small
        \SetKwInOut{Input}{Input}
        \SetKwInOut{Output}{Output}
        \SetKwProg{Fn}{Function}{ is}{end}
        \caption{Vectorized Jacobi's method}\label{Algo:vectorized jacobi}
        \Input{symmetric matrix $A \in \mathbb{R}^{d \times d}$}
        \Output{The eigenvalues and eigenvectors of input $A$}
        \BlankLine
        $M \leftarrow \lfloor \frac{d+1}{2} \rfloor$ \;
        \While{True}{
          $i \leftarrow 0$ \;
          $J_{i} \leftarrow I_{d}$ \;
          $Q \leftarrow I_{d}$ \;
          \For{i < 2*M}{
            $k\_list, l\_list \leftarrow ElementSelection(i)$ \;
            $tar\_elements \leftarrow A[k\_list, l\_list]$ \;
            $tar\_diff \leftarrow A[k\_list, k\_list] - A[l\_list, l\_list]$ \;
            $\tau\_list \leftarrow tar\_diff / (2*tar\_elements^{2}$) \;
            $flag\_zero \leftarrow tar\_elements == 0$\;
            $tan\_list \leftarrow |\tau\_list| + \sqrt{({\tau\_list}^2 + 1)}$ \;
            $tan\_list \leftarrow (1-flag\_zero)*1/tan\_list$ \;
            $cos\_list \leftarrow 1/(\sqrt{tan\_list^2 + 1})$ \;
            $sin\_list \leftarrow tan\_list * cos\_list$ \;
            \BlankLine
            $J_{i}[k\_list, k\_list], J_{i}[l\_list, l\_list] \leftarrow cos\_list$ \;
            $J_{i}[k\_list, l\_list] \leftarrow -sin\_list$ \;
            $J_{i}[l\_list, k\_list] \leftarrow sin\_list$ \;
            $A \leftarrow J_{i}^{T}AJ_{i}$ \;
            $Q \leftarrow J_{i}^{T}Q$ \;
            $i += 1$\;
            \tcc{Check for convergence every $r$ rounds.}
            \If{Reveal($mean(A^{k,l}_{k\neq l}) < threshold$)}{
              \Return{$diag(A), Q$};
            }
          }
        }
      \end{algorithm}

We iterate until reaching the convergence threshold, $1e-5$.  The threshold ensures each of the calculated eigenvalue $\lambda_{i}$ and eigenvector $v_{i}$ can achieve the accuracy of three significant numbers for $Av_{i} == \lambda_{i} v_{i}, 0 < i < d$ for the input $A$. The reason for choosing the mean value as the stoping criterion is to avoid accumulating the accuracy error in cipher-text. On line 23, in order to check whether the algorithm has converged,  we reveal the comparison result (0 or 1) to the computation servers running MPC protocol every $r$ iterations.  The servers only gain a bit sequence $s = \{0, 0, ..., 0, 1\}$, revealing the length of $s$, $|s|$.  The length $|s|$ only roughly reveals the dimensions of the covariance matrix, which is already known to the servers.  Thus, we do not think it is a privacy risk. 




\para{Performance Comparison between QR and Jacobi.  }
The main computation cost for both algorithms in cipher-text comes from the total numbers of the \emph{expensive operations (EOs)} based on the analysis in Section~\ref{section:MPC_features}. The total number of EOs in QR shift and Jacobi is mainly determined by the per-iteration EO count and the number of iterations to convergence.





In each HR in QR shift, there are four EOs, including two \code{sqrt}, one \code{reciprocal}, and one \code{comparison}. In each iteration of QR shift, we need $(d-1)$ HRs ($d$ is the dimension of the covariance matrix), with vector dimension changing from $d$ to $1$ sequentially.  In each GR in Jacobi, there are at least three \code{reciprocal}, one \code{comparison}, one \code{sqrt}, and an additional \code{equal} to avoid the data overflow in MPC.  
The first two rows of Table~\ref{Table:comp for two transformations} summarize EOs per iteration in each algorithm. 


We explore the convergence iterations empirically and Section~\ref{sec:eval_analysis} summarizes the reaults. From the results, we observe that with the setting of our threshold, $1e-5$, the convergence iterations of QR shift is $O(d^{\alpha})$ where $0.74 < \alpha < 0.80$, combining with the $(d-1)$ HR per iteration, the total number of EOs is around $O(d^{\alpha + 1})$ which is higher than $O(d)$ on Jacobi. We want to point out that although Jacobi takes more iterations to converge, it is able to batch up more operations in each iteration. The overall performance is actually faster than QR shift.

\begin{table}[tb]
  \small
  \caption{EOs for Each Iteration}\label{Table:comp for two transformations}
  \begin{tabular}{lccccl}
    \toprule
      & \code{comp} & \code{eq} & \code{sqrt} & \code{reciprocal} \\
    \midrule
    HR in QR Shift & $1\times(d-1)$ & 0 & $2\times(d-1)$ & $1\times(d-1)$\\
    GR in Jacobi & 1 & 1 & 2 & 3\\ 
    Optimized Jacobi & 2 & 1 & 2 & 1 \\
  \bottomrule
\end{tabular}
\end{table}

\para{EO-Reduction for Jacobi.  }\label{section:algorithm optimization}
We observe that we can further optimize Jacobi by replacing more expensive EOs with cheaper ones.  As we discussed, \code{reciprocal} is over $10\times$ more expensive than \code{comparison} due to the numerical algorithm.  We replace line 10 to 15 with the following code in Algorithm~\ref{Algo:Transformed vectorization Jacobi's method}.  The resulting algorithm reduces \code{reciprocal} per iteration from three to one, at the cost of one extra \code{comparison}.  The last row in Table~\ref{Table:comp for two transformations} summarizes the EOs in this algorithm.  

\begin{algorithm}[tb]
	\small
	\renewcommand{\algorithmcfname}{Algorithm}
	\caption{Transformed Rotation Calculation}\label{Algo:Transformed vectorization Jacobi's method}
	$\cos2\theta\_list \leftarrow |tar\_diff|/(\sqrt{4*tar\_elements^{2} + tar\_diff^{2}})$ \;
	$\cos^{2}\_list \leftarrow 0.5 + 0.5*\cos2\theta\_list$ \;
	$\sin^{2}\_list \leftarrow 0.5 - 0.5*\cos2\theta\_list$ \;
	$\theta\_list \leftarrow \sqrt{[\cos^{2}\_list, sin^{2}\_list]}$\;
	$\cos\_list \leftarrow \theta\_list[0]*(1-flag\_zero) + flag\_zero$ \;
	$\sin\_list \leftarrow \theta\_list[1]*((tar\_elements*tar\_diff > 0)*2-1)$;
\end{algorithm}

In line 4 in Algorithm~\ref{Algo:Transformed vectorization Jacobi's method}, we concatenate $\cos{2\theta}$ and $\sin{2\theta}$ vector to perform the \code{sqrt} operation as a single vector operation.  Then on Line 5 and 6, we separate the $\cos{\theta}$ and $\sin{\theta}$.  This is beneficial given the observations of batching up in Table~\ref{Table:time costs}.

\subsection{Projection Matrix and Inference}
After the eigen-decomposition step, we sort all the eigenvalues to select the largest $K$ eigenvalues and the corresponding eigenvectors to construct the projection matrix $P$.  The key step is to sort the eigenvalues for selection.  As sorting requires many relatively expensive \code{comparison}'s, we want to batch up as much as possible. Thus, we design the \code{batch\_sort} algorithm that combines $O(d^2)$ comparisons into a single vector comparison of size $d^2$.  Note that it is a tradeoff between memory space and batch size.  Using $O(d^2)$ memory, we can greatly accelerate the comparison just by batching up.  Algorithm~\ref{Algo:batch sort} shows the \code{batch\_sort} algorithm.  


 
  \begin{algorithm}[t]
    \small
    \SetKwInOut{Input}{Input}
    \SetKwInOut{Output}{Output}
    \SetKwProg{Fn}{Function}{ is}{end}
    \caption{Batch sort algorithm}\label{Algo:batch sort}
    \Input{one-dimension column vector $x = (x_{0}, x_{1}, ..., x_{n-1})^T$}
    \Output{The sorted array $x$ with the \code{argsort} result $index$}
    \BlankLine
    $X \leftarrow repeat(x, n)$ \;
    $C \leftarrow X^{T} < X$ \tcc*[r]{Comparision matrix}
    $E \leftarrow (X^{T} == X)$ \tcc*[r]{Equal matrix}
    $E \leftarrow (cumsum(E, axis=0) == E) \times E$ \;
    $B \leftarrow (cumsum(E, axis=0) - E) \times E$ \;
    $d \leftarrow sum(C, axis=1) + sum(B, axis=1)$ \tcc*[r]{Sorted index}
    $D \leftarrow repeat(d, n)$ \;
    $r \leftarrow range(0, n)^T$                   \tcc*[r]{Range}
    $R \leftarrow repeat(r, n)$ \;
    $M \leftarrow (D^{T} == R) $                   \tcc*[r]{Mask matrix}
    \Return{$sum(X^{T} \times M, axis=1)$, d}
  \end{algorithm}

We use matrix $E$ in Algorithm~\ref{Algo:batch sort} to handle elements with the same value. Lines 4 and 5 update the indices for each same-value element sequentially. Finally, we return the sorted array with its corresponding index in cipher-text. Note that during the entire process, we leak no information about the elements. All the position updates are achieved with the comparison result as an indicator array in cipher-text. We re-arrange each element with the result of \code{argsort} based on the MPS protocol in Section~\ref{section:MPC protocols}. Then we can take the largest $K$ eigenvalues and the corresponding eigenvectors to form the projection matrix $P \in \mathbb{R}^{d \times K}$.


After computing the projection matrix $P$, we can perform a series of downstream tasks such as dimension reduction and anomaly detection, without decrypting $P$. The dimension reduction is a matrix multiplication of $X \in \mathbb{R}^{n \times d}$ and the projection matrix $P$, which can be done in MPC fast.  

\section{Evaluation}

We evaluate our design in two aspects, performance and effectiveness.  In Section~\ref{sec:eval_overview} and~\ref{sec:eval_analysis}, we use several open datasets and synthetic data to demonstrate the performance of each step of the PCA process.  We also use the real-world datasets to evaluate the effects when integrating the data from multiple parties. 

We conduct all evaluations on a four-server PrivPy ~\cite{li2019privpy} deployment.  Four independent servers are the minimal configuration for PrivPy's \textit{(4,2)-secret sharing} scheme.  All servers contain two 20-core 2.0 GHz Intel Xeon processors. 

 
We have used several datasets for performance evaluation.  The first two columns in Table~\ref{Table:benchmark} summarizes the size of the datasets.  We choose the Wine dataset~\cite{cortez2009modeling} because it is used in related work~\cite{liu2020privacy} and others because they represent different data sizes and workloads for dimension reduction. The details of IoT~\cite{meidan2018n} and MOOC~\cite{yu2020mooccube} datasets's multi-party construation are in the followings. On each of the dataset, we compared the \textit{explained variance ratio (EVR)} with plain-text PCA implementation in Scikit-learn~\cite{pedregosa2011scikit} of the computed principal components $\lambda_i$, where $EVR(\lambda_{i}) = \frac{\lambda_i}{\sum_{j=1}^{d}{\lambda_j}}, 0<i\leq10$, the mean precision error is within $1e-3$, thus confirming the correctness of our implementation.

\begin{table}[tb]
	\small
	\caption{Datasets in Evaluation and their running time (sec)}
	\label{Table:benchmark}
	\begin{tabular}{lcccccl}
		\toprule
		Name & Size & Cov & Decomp & Sort & Inference \\
		\midrule
		Wine~\cite{cortez2009modeling} & $6,497 \times 11$ & 0.56 & 2.27 & 0.04 & 0.01 \\
		Insurance~\cite{meng2005comprehensive} & $9,822 \times 85$ & 2.97 & 56.05 & 0.11 & 0.07 \\
		Musk~\cite{aggarwal2015theoretical} & $3,062 \times 166$  & 3.29 & 300.76 & 0.29 &  0.12 \\
		IoT-3~\cite{meidan2018n} & $ 3,803,677 \times 115$ & 3.59 & 136.50 & 0.16 & 0.01 \\
		IoT-5~\cite{meidan2018n} & $ 5,866,616 \times 115$ & 3.62 & 142.98 & 0.15 & 0.01 \\
		IoT-9~\cite{meidan2018n} & $7,062,606 \times 115$ & 3.64 & 142.80 & 0.17 & 0.01 \\
		MOOC-3~\cite{yu2020mooccube} & $195,177 \times 60$ & 13.26 & 44.59 & 0.10 & 0.01\\
		MOOC-7~\cite{yu2020mooccube} & $195,177 \times 140$ & 31.55 & 295.03 & 0.23 & 0.02\\
		MOOC-10~\cite{yu2020mooccube} & $195,177 \times 200$ & 50.20 & 808.53 & 0.40 & 0.03 \\
		\bottomrule
	\end{tabular}
\end{table}

\subsection{Performance Overview}
\label{sec:eval_overview}

\para{Micro-benchmarks.  }
The first three rows of Table~\ref{Table:benchmark} summarizes the results of performance micro-benchmarks.  In these benchmarks, we omit the time to send the data from multiple parties and just focus on the computation time.  We report time cost in the four steps.  The Cov step includes the time to encrypt the entire dataset and perform operations to compute the covariance matrix.  We have the following observations:

1) Our method achieves reasonable time for the PCA task.  In fact, our method is much faster than previously reported results.  
Comparing to \cite{al2017privacy} that does the decomposition in full cipher-text and takes 126.7 minutes on a $50 \times 50$ matrix, our method achieves a $200\times$ speed-up on matrix with a similar scale. 
Even comparing to ~\cite{liu2020privacy} that reveals the covariance matrix as plain-text, takes around three seconds on the same Wine dataset, and we can do everything in cipher-text within 5 seconds. 

2) Matrix decomposition is the most time-consuming step.  With the dimension of feature space $d$ increasing, the time for matrix decomposition is between $O(d^2)$ and $O(d^3)$. The reason for this cost is higher than $O(d^2)$ is because iterations for convergence is also at least linear to $d$, which we will discuss in the next section.  However, as we can aggressively batch up operations in Jacobi, we obtain a final execution time less than $O(d^3)$.

\para{Horizontally-partitioned IoT dataset.  }
We use a larger scale dataset~\cite{meidan2018n} with a dimension of $7,062,606 \times 115$ to evaluate the scalability on horizontally partitioned dataset.  We emulate horizontal partition by letting each IoT device be a separate party, and we need to allow each party to preprocess the data before combining them locally (Eq.~\ref{eq:horizontal-preprocess} in Section~\ref{section:hp}).  Row 4 to 6 in Table~\ref{Table:benchmark} shows the performance. The preprocessing time includes a local preprocess time (same as discussed previously) of 0.76 seconds and a combination time on cipher-text of around 2.8 seconds.

\para{Vertically-partitioned MOOCCube dataset.  }\label{exp:mooc}
We use a derived dataset from MOOCCube~\cite{yu2020mooccube} as an example of a vertically partitioned dataset.  Each row represents a student (identified by an integer ID) and each column represents a concept (i.e., a topic in the course, such as ``binary tree'' in a programming course).  We choose $200$ concepts and separate them into ten courses. Treating these courses are from different institutions that cannot share the data, but we want to use the student's learning record on all 200 concepts for analysis.  Thus, it becomes a vertically partitioned situation.  There are in total $195,177$ unique student IDs, and thus each party has a dataset with size $195,177 \times 20$.  

The last three rows in Table~\ref{Table:benchmark} shows the performance. The Cov time contains the local arrangement which is around $0.21$ seconds and the remaining is the combination time using MPC. The entire preprocessing time is much higher than other experiments as we have to encrypt all the samples from vertical partitions. Note that as more parties result in more dimensions, both decomposition and the preprocessing time increase with the number of vertically partitioned parties, as expected. 


\subsection{Analysis of Optimizations}
\label{sec:eval_analysis}

  \begin{figure}
    \centering
    \subfigure[Iterations for QR shift]{
      \includegraphics[scale=0.256]{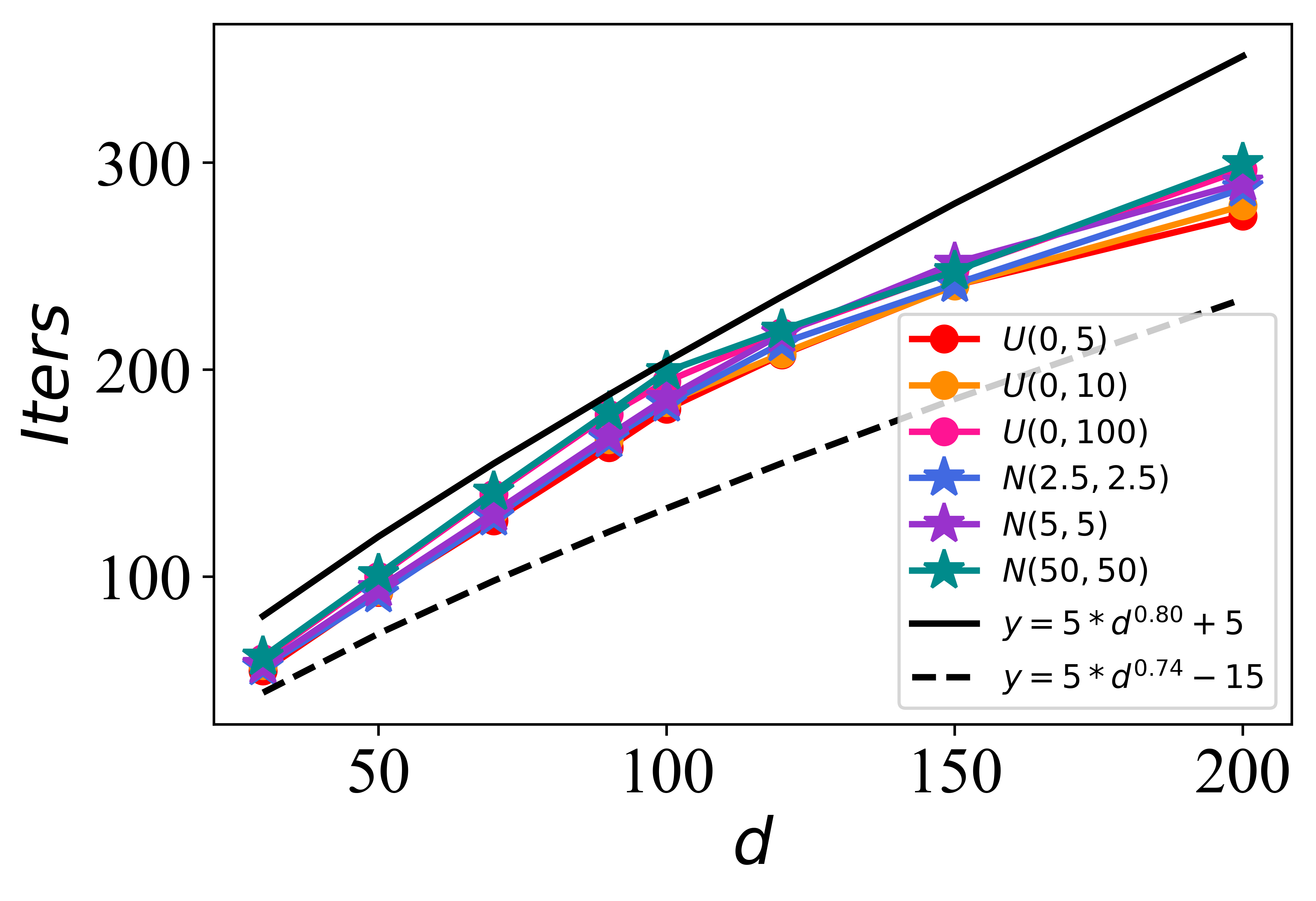}\label{fig:conv_qr}
    }
    \subfigure[Iterations for Parallel Jacobi]{
      \includegraphics[scale=0.256]{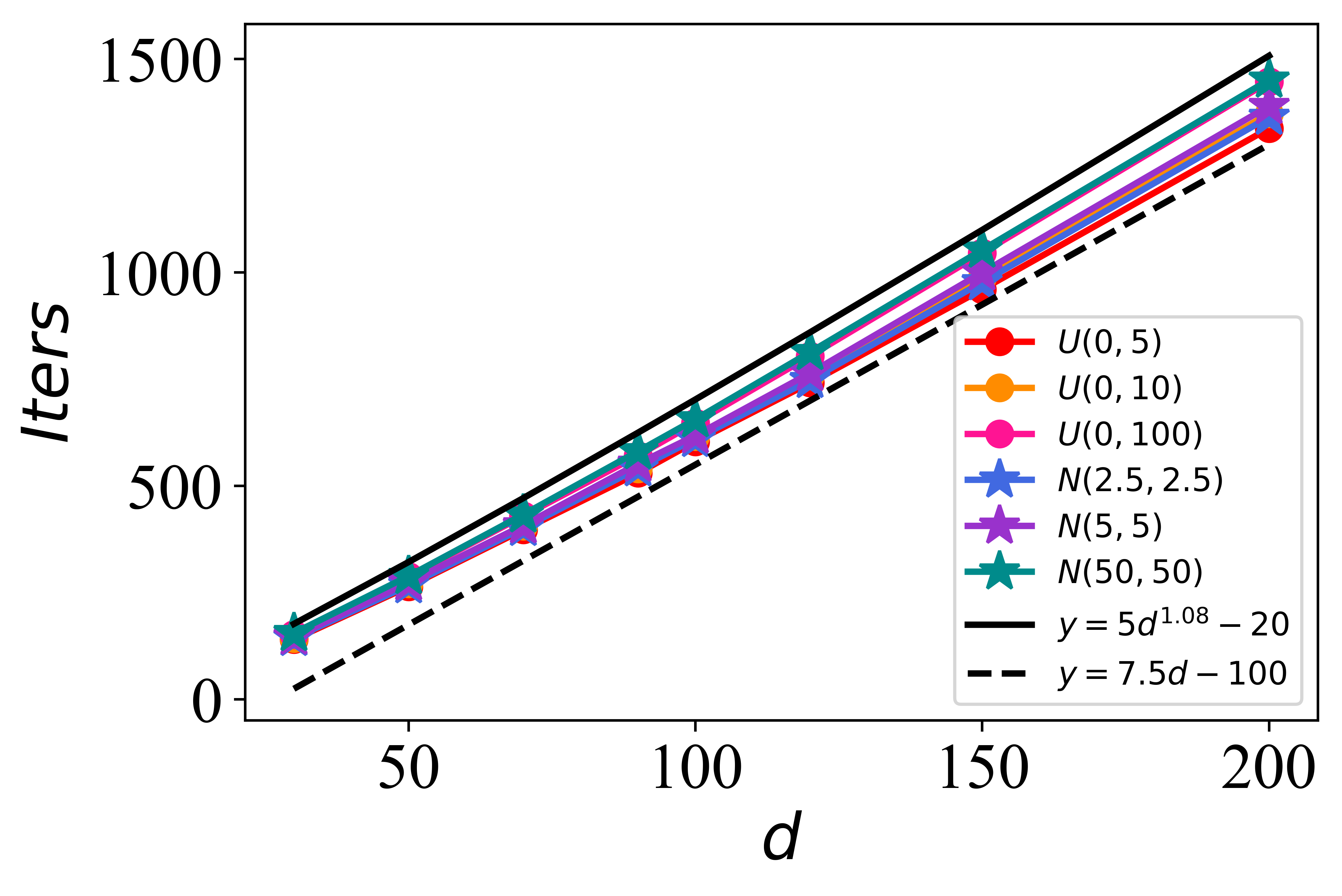}\label{fig:conv_jacobi}
    }
    \caption{Convergence Illustration}\label{fig:convergence}
  \end{figure}


\para{Iterations to converge } is a very important metric affecting performance.  We experimentally evaluate the convergence iterations with different dimensions as well as data distributions using synthetic datasets for both 
QR shift and Jacobi repeating 50 times each with different random seeds. Figure~\ref{fig:convergence} shows the results.  We observe that 
1) For both algorithms, the number of iteration mainly depends on the matrix dimension $d$, where QR shift is about $O(d^{\alpha})$ where $0.74 < \alpha < 0.8$ and Jacobi is about $O(d)$.   
2) The data distribution can affect the convergence at some level.  One reason causing this phenomenon is because our threshold in Algorithm~\ref{Algo:vectorized jacobi} is relatively looser than the plain-text version to avoid accumulating the accuracy error. Thus matrices with smaller off-diagonal elements converge earlier than those with larger elements.

\para{Jacobi's method vs. QR shift.  }
The first two rows in Table\ref{Table:time compare} shows the comparison between QR Shift and Jacobi on cipher-text. This is because each iteration of QR uses $(d-1)$ HRs that need to run sequentially without the benefit of batching up, which is consistent with the analysis in Section~\ref{section:md}.



   
    \begin{table}[tb]
      \small
      \centering
      \caption{Time consumption(s) comparison}\label{Table:time compare}
      \begin{tabular}{ccccc}
        \toprule
         & 50 & 70 & 100 & 120\\
        \midrule
        QR Shift & 325.49 & 648.03 & 1367.63 & 1981.08 \\
        Jacobi & 48.53 & 95.51 & 180.17 & 253.35\\
        Jacobi w/ EO-reduction & 36.01 & 71.63 & 148.14 & 215.82\\
        \bottomrule
      \end{tabular}
    \end{table}


\para{Benefits of EO Reduction.  } 
We have compared the time for the same matrix decomposition before and after the EO reduction in the last two rows of Table~\ref{Table:time compare}.  Replacing 2 \code{reciprocal} with one \code{comp} in each iteration, we show that we can obtain a $18\%$ performance gain.
  

\subsection{Benefits of Data Integration}\label{Section:exp-effect}

Back to the motivation why we need computing PCA with multi-party datasets, using the horizontally partitioned IoT~\cite{meidan2018n} and vertically partitioned MOOC~\cite{yu2020mooccube} datasets, we show that joint PCA does improve downstream task performance.  



	\begin{table}
		\small
		\centering
		\caption{Benefits for data integration}\label{Table:effects}
		\begin{tabular}{c|cccc|ccccl}
			\toprule
			 & \multicolumn{4}{c}{IoT} & \multicolumn{4}{c}{MOOC} \\
			~ & N=1 &  N=3 &  N=5 &  N=9 & N=1 & N=3 & N=7 & N=10 \\
			\midrule
			Precision & 0.79 & 0.80 & 0.99 & 1.0 & 0.74 & 0.79 & 0.80 & 0.87 \\
			Recall & 0.74 & 0.92 & 0.98 & 1.0 & 0.81 & 0.82 & 0.83 & 0.83 \\
			f1-score & 0.71 & 0.89 & 0.99 & 1.0 & 0.73 & 0.74 & 0.77 & 0.77 \\
			\bottomrule
		\end{tabular}
	\end{table}

We perform a similar task on both datasets:  first, we use our privacy-preserving PCA method to reduce the feature dimension to $K$.  Then we train a classifier to perform the classification task on the $K$-dimension feature matrix and compare the precision, recall and F1-score.  Table~\ref{Table:effects} summarizes results on both datasets.  

\para{Horizontally partitioned IoT dataset. }\label{exp: IoT fuse}
Recall that the dataset is partitioned into 9 different IoT devices, each with the same 115 feature dimensions. The classification task is to determine whether the device is under attack by botnet \emph{gafgyt}, \emph{miari} or not, the hyperparameter of $K = 20$ and the classifier is \emph{AdaBoost}. Table~\ref{Table:effects} shows that integrating data from more parties (i.e., device types) significantly boosts the downstream classification performance, because actually many attacks happen on either \textit{Ennio\_doorbell} or \textit{Samsung\_Webcam}, without using data from these parties, the PCA algorithm fails to pickup feature dimensions required to capture the \emph{miari} botnet. 


  

\para{Vertically partitioned MOOC dataset. }\label{exp: MOOCube fuse}
We show the effects of vertically partitioned data integration using the MOOCCube dataset\cite{yu2020mooccube}. Each course constructs the feature space with 20 dimensions using assigned concepts. The prediction task we constructed is to predict whether a student will enroll in some courses (top 400 who has more relations with the selected concepts). The hyperparameter $K = 7$ and the classifier here is \emph{GradientBoosting}. Table~\ref{Table:effects} shows that with related features expanded, the model can gain better performance.

\section{Conclusion and Future Work}
\label{sec:conclusion}

Privacy has become a major concern in data mining, and MPC seems to provide a technically sound solution to the privacy problem.  However, existing \emph{basic-operation-level} optimizations in MPC still do not provide sufficient performance for complex algorithms.  Using PCA as an example, we are among the first work to show that by carefully choosing the algorithm (e.g., Jacobi vs. QR), replacing individual operations based on the MPC performance characteristics (e.g. replacing \code{reciprocal} with \code{comparison}), and batching up as much as possible (e.g., \code{batch\_sort}), we can provide an \emph{algorithm-level} performance boost, running 200$\times$ faster over existing work with similar privacy guarantee.  

As future work, we will verify our approach on more MPC platforms, especially those with different basic operator performance, and expand the methodology to other data mining algorithms.  


\bibliographystyle{ACM-Reference-Format}
\bibliography{ms}

\newpage
\appendix
\section{Privacy-preserving Sqrt and Reciprocal Operations}\label{sqrt and reciprocal}

  The calculation for sqrt and reciprocal is based on \textit{Newton-Raphson} iterations, here we have designed some transformations to accelerate the convergence rate.

  \subsection{Privacy-preserving Sqrt} 
  The square root of value $A$ can be estimated through the solution of function $f(x) = \frac{1}{x^2} - A = 0$, where the solution $x \rightarrow \frac{1}{\sqrt{A}}$. The iteration of $x$ was:
    \begin{equation}
      \begin{aligned}
        & x_{0} \leftarrow appro, \\
        & x_{k+1} = \frac{1}{2}x_{k}(3 - Ax_{k}^2),\\
      \end{aligned}
    \end{equation}

  The $appro$ we choose here equals $0.48$. It is estimated through the expection of $\frac{1}{\sqrt{A}}$ for $A \sim U [1, 10]$ while less than $\sqrt{\frac{3}{A}}$ which the convergence condition. In our algorithm, we will scale the $appro$ for input value $A$ not in $[1, 10]$. From experiments, we can get convergence with around $7$ itertaions with less than $1e-14$ errors.

  \subsection{Privacy-preserving Reciprocal} 
  The reciprocal of input value $A$ can be estimated through the solution of function $f(x) = \frac{1}{x} - A = 0$. Here we using the second-order \textit{Newton-Raphson} approximation with iterations reduced. The update equation of $x$ was:
    \begin{equation}
      \begin{aligned}
        & x_{0} \leftarrow appro, \\
        & x_{k+1} = x_{k} + x_{k}(1-Ax_{k}) + x_{k}{(1-Ax_{k})}^2, \\
      \end{aligned}
    \end{equation}
    The initial value is defined through a wide range of possible values. We will firstly compare the input value $A$ with vector $\mathbf{a} = [10^{-11}, ... 1, ..., 10^{12}]$ and select the nearest estimation reciprocal in $[10^{11}, ... 1, ..., 10^{-12}]$ with the encrypted $\{0, 1\}$ arrays through \textit{Oblivious Transfer} protocol. From experiments, we can get convergence with around $6$ iterations ess than $1e-14$ errors.

\section{orthogonal Transformations}\label{orthogonal transformation}
  Here we introduce the details of two orthogonal transformations where we can see the number of EOs each for Section~\ref{section:md}. The \code{sign} function requires one time \code{comp}.
  \begin{algorithm}[h]
    \small
    \renewcommand{\algorithmcfname}{Orthogonal Transformation}
    \SetKwInOut{Input}{Input}
    \SetKwInOut{Output}{Output}
    \SetKwProg{Fn}{Function}{ is}{end}
    \caption{Householder Reflection}
    \Input{vector $w \in \mathbb{R}^{n}$}
    \Output{Orthogonal matrix $P \in \mathbb{R}^{n\times n}$ where $Pw = ke$, $k = ||w||_{2}$}
    \BlankLine
    \Fn{Householder(w) : P}{
      $u \leftarrow w$ \;
      $u[1] \leftarrow u[1] + sign(u[1])||u||_{2}$ \;
      $u \leftarrow u/||u||_{2} $ \;
      $P \leftarrow I_{n} - 2*uu^{T}$ \;
      \Return{P}
    }
  \end{algorithm}  
  \begin{algorithm}[h]
    \small
    \renewcommand{\algorithmcfname}{Orthogonal Transformation}
    \SetKwInOut{Input}{Input}
    \SetKwInOut{Output}{Output}
    \SetKwProg{Fn}{Function}{ is}{end}
    \caption{Givens Rotation}
    \Input{symmetric matrix $A \in \mathbb{R}^{n \times n}$ and index $k, l$}
    \Output{Orthogonal matrix $J_{kl} \in \mathbb{R}^{n\times n}$ which can turn elements $a_{kl}$ and $a_{l,k}$ to zero}
    \BlankLine
    \Fn{GivensRotate(A, k, l) : $J_{kl}$}{
      $J_{kl} \leftarrow I_{n}$ \;
      $\tau \leftarrow (A[k, k] - A[l, l])/(2*A[k, l])$ \;
      $t \leftarrow sign(\tau)/(|\tau| + \sqrt{1 + \tau^2})$ \;
      $c \leftarrow 1/\sqrt{1+\tau^2}$ \;
      $s \leftarrow tc$ \;
      $J_{kl}[k, l] \leftarrow -s; J_{kl}[l,k] \leftarrow s$ \;
      $J_{kl}[k, k], J_{kl}[l, l] \leftarrow c$ \;
      \Return{$J_{kl}$}
    }
  \end{algorithm}

\section{Privacy-preserving QR Shift Algorithm}\label{pp QR shift}
  The implementation for our cipher-text QR shift algorithm including two phase and based on householder reflection, first is to reduce the origional $N\times N$symmetric matrix to tradtional form implementing thourgh $N-2$ step householder reflections; second is to reduce the tridiagonal matrix into diagonal using QR decomposition with Rayleigh shift.
  \begin{algorithm}
    \small
    \SetKwInOut{Input}{Input}
    \SetKwInOut{Output}{Output}
    \caption{QR shift with Rayleigh quotient}
      \Input{$N\times N$ symmetric encrypted matrix $A$}
      \Output{Upper tradtional $T$ with $Q$ such that $A = Q^{T}TQ$ where $Q_{0}$ is an orthogonal matrix}
      \BlankLine
      $A_{0} \leftarrow A$ ;
      $Q \leftarrow I_{N}$ \;
      \For{$i \leftarrow 0$ \KwTo $N-2$}{
        $u_{i} \leftarrow A_{i}[i+1:N, k]$ \;
        $P_{i} = I_{N}$\; 
        $\mathbf{P_{i}[i+1:N, i+1:N] = Householder(u_{i})}$

        $Q[i+1:N, i+1:N] \leftarrow P_{i}Q[i+1:N, i+1:N]$\;
        $A_{i+1} \leftarrow P_{i}A_{i}P_{i}^{T}$ \;
      }
      $T \leftarrow A_{N-1}$ \;
  \end{algorithm}

  \begin{algorithm}
    \small
    \SetKwInOut{Input}{Input}
    \SetKwInOut{Output}{Output}
    \SetKwProg{Fn}{Function}{ is}{end}
    \caption{QR shift - Second Phase: Rayleigh quotient shift QR decomposition}
    \Input{Tridiagonal matrix $T$ and orthogonal $Q$ returned from algorithm1}
    \Output{eigenvalues and eigenvectors of $T$ which is the same as $A$}
    \Fn{QR(A) : Q, R where A = QR}{
      $Q \leftarrow I_{N}$ \;
      $R \leftarrow copy(A)$ \;
      \For{$i \leftarrow 0$ \KwTo $N-1$}{
        $u_{i} \leftarrow R[i:, i]$ \;
        $P_{i} = I_{N}$\; 
        $\mathbf{P_{i}[i:N, i:N] = Householder(u_{i})}$ \;
        $R \leftarrow P_{i}R$ \;
        $Q \leftarrow QP_{i}$ \;
      }
    }
  \end{algorithm}

  \section{Privacy-preserving SVD based on vectorized Jacobi's method}\label{ppsvd}

    \begin{algorithm}
      \small
      \SetKwInOut{Input}{Input}
      \SetKwInOut{Output}{Output}
      \SetKwProg{Fn}{Function}{ is}{end}
      \caption{Privacy-preserving vectirized SVD}\label{Algo:vectorized svd}
      \Input{real-value matrix $G \in \mathbb{R}^{m \times n}$}
      \Output{The singular values $\Sigma$ with the left singular vector matrix $U$ and the right singular vector matrix $V$}
      \BlankLine
      $M \leftarrow \lfloor \frac{n+1}{2} \rfloor$ \;
      \While{True}{
        $i \leftarrow 0$, $J_{i} \leftarrow I_{n}$  \;
        $Q \leftarrow I_{n}$, $A \leftarrow G^{T}G$\;
        \For{i < 2*M}{
          $k\_list, l\_list \leftarrow ElementsSelection(i)$ \;
          $tar\_elements \leftarrow A[k\_list, l\_list]$ \;
          $tar\_diff \leftarrow A[k\_list, k\_list] - A[l\_list, l\_list]$ \;
          $\cos2\theta\_list \leftarrow |tar\_diff|/(\sqrt{4*tar\_elements^{2} + tar\_diff^{2}})$ \;
          $\cos^{2}\_list \leftarrow 0.5 + 0.5*\cos2\theta\_list$ \;
          $\sin^{2}\_list \leftarrow 0.5 - 0.5*\cos2\theta\_list$ \;
          $\theta\_list \leftarrow \sqrt{[\cos^{2}\_list, sin^{2}\_list]}$\;
          $\cos\_list \leftarrow \theta\_list[0]*(1-flag\_zero) + flag\_zero$ \;
          $\sin\_list \leftarrow \theta\_list[1]*((tar\_elements*tar\_diff > 0)*2-1)$\;

          \BlankLine
          $J_{i}[k\_list, k\_list], J_{i}[l\_list, l\_list] \leftarrow cos\_list$ \;
          $J_{i}[k\_list, l\_list] \leftarrow -sin\_list$ \;
          $J_{i}[l\_list, k\_list] \leftarrow sin\_list$ \;
          $G \leftarrow GJ_{i}$ \;
          $Q \leftarrow QJ_{i}$ \tcc*[r]{for the right singular vector}
          $i += 1$ \;

          \tcc{Check for convergence every $r$ rounds}
          \If{Reveal($mean(A^{k,l}_{k\neq l}) < threshold$)}{
            $\Sigma \leftarrow [\sigma_{1}, ..., \sigma_{n}]\, where\, \sigma_{i} = ||G[:, i]||_{2}$ \;
            $U \leftarrow [u_1, ... u_n] \, where \, u_i = G[:, i]/\sigma_{i}$ \;
            $V \leftarrow Q$ \;
            \Return{$\Sigma, U, V$}
          }
        }
      }
    \end{algorithm}
    Just as we mentioned in section\ref{section:md}, based on \textit{One-Sided Jacobi}, we can using the same optimization to implement the privacy-preserving SVD, showin in Algorithm~\ref{Algo:vectorized svd}.

\end{document}